\def\Version{ 2.5             
    }







\message{ Assuming 8.5" x 11" paper }    

\magnification=\magstep1	          

\raggedbottom

\parskip=9pt

%

\def\singlespace{\baselineskip=12pt}      
\def\sesquispace{\baselineskip=16pt}      




\font\openface=msbm10 at10pt
 %

\def\Reals         {{\hbox{\openface R}}}
\def\Complexes     {{\hbox{\openface C}}}
\def\alfa{\alpha}
\def\SetOf#1#2{\left\{ #1  \,|\, #2 \right\} }
\def\isom{\simeq}		
\def\implies{\Rightarrow}

\def\braces#1{ \{ #1 \} }
\def\compose{\circ}
\def\tensor{\otimes}		


%


\let\miguu=\footnote
\def\footnote#1#2{{$\,$\parindent=9pt\baselineskip=13pt%
\miguu{#1}{#2\vskip -7truept}}}
 %

\def\linebreak{\hfil\break}
\def\lbr{\linebreak}
\def\pagebreak{\vfil\break}


\def\BulletItem #1 {\item{$\bullet$}{#1 }}
\def\bulletitem #1 {\BulletItem{#1}}

%
%




\def\REMARK{\noindent {\csmc Remark \ }}
\def\DEFINITION{\noindent {\csmc Definition \ }}

\def\PROOF{\noindent \quad{\csmc Proof \ }}

\def\LEMMAnumbered#1{\smallskip\noindent {\csmc Lemma #1.}}

\def\PrintVersionNumber{
 \vskip -1 true in \medskip 
 \rightline{version \Version} 
 \vskip 0.3 true in \bigskip \bigskip}

\def\author#1 {\medskip\centerline{\it #1}\bigskip}

\def\address#1{\centerline{\it #1}\smallskip}

\def\furtheraddress#1{\centerline{\it and}\smallskip\centerline{\it #1}\smallskip}

\def\email#1{\smallskip\centerline{\it address for email: #1}} 

\def\AbstractBegins
{
 \singlespace                                        
 \bigskip\leftskip=1.5truecm\rightskip=1.5truecm     
 \centerline{\bf Abstract}
 \smallskip
 \noindent	
 } 
\def\AbstractEnds
{
 \bigskip\leftskip=0truecm\rightskip=0truecm       
 }

\def\section #1 {\bigskip\noindent{\headingfont #1 }\par\nobreak\smallskip\noindent}

\def\subsection #1 {\medskip\noindent{\subheadfont #1 }\par\nobreak\smallskip\noindent}
 %

\def\ReferencesBegin
{
 \singlespace					   
 \vskip 0.5truein
 \centerline           {\bf References}
 \par\nobreak
 \medskip
 \noindent
 \parindent=2pt
 \parskip=6pt			
 }
 %

\def\reference{\hangindent=1pc\hangafter=1} 

\def\ref{\reference}

\def\sepref{\parskip=8pt \par \hangindent=1pc\hangafter=0}
 %

\def\journaldata#1#2#3#4{{\it #1\/}\phantom{--}{\bf #2$\,$:} $\!$#3 (#4)}
 %

\def\eprint#1{{\tt #1}}

 %

 %


\def\webhome{{\tt http://www.pitp.ca/personal/rsorkin/}}
 %

 %



\def\webtilde{\lower2pt\hbox{${\widetilde{\phantom{m}}}$}}

 %

\def\hpf#1{\webhome{\tt{some.papers/}}}
 %

\def\hpfll#1{\webhome{\tt{lisp.library/}}}
 %


\font\titlefont=cmb10 scaled\magstep2 

\font\headingfont=cmb10 at 12pt
 %

\font\subheadfont=cmssi10 scaled\magstep1 
 %

\font\csmc=cmcsc10  







\font\german=eufm10 at 10pt
\def\Buchstabe#1{{\hbox{\german #1}}}
\def\S{\Buchstabe{S}}
\def\T{\Buchstabe{T}}
\def\H{\Buchstabe{H}}
\def\A{\Buchstabe{A}}

\def\M{{\cal{M}}}

\def\VEC{\hbox{VEC}}
\def\TVEC{\hbox{TVEC}}

\def\P{{\bf {P}}}
\def\fhat{\hat{f}}


\newcount\tmpnum \newdimen\tmpdim
{\lccode`\?=`\p \lccode`\!=`\t  \lowercase{\gdef\ignorept#1?!{#1}}}

\edef\widecharS{\expandafter\ignorept\the\fontdimen1\textfont1}

\def\widebar#1{\futurelet\next\widebarA#1\widebarA}
\def\widebarA#1\widebarA{%
   \def\tmp{0}\ifcat\noexpand\next A\def\tmp{1}\fi
   \widebarE
   \ifdim\tmp pt=0pt \overline{#1}%
   \else {\mathpalette\widebarB{#1}}\fi
}
\def\widebarB#1#2{%
   \setbox0=\hbox{$#1\overline{#2}$}%
   \tmpdim=\tmp\ht0 \advance\tmpdim by-.4pt
   \tmpdim=\widecharS\tmpdim
   \kern\tmpdim\overline{\kern-\tmpdim#2}%
}

\def\widebarC#1#2 {\ifx#1\end \else 
   \ifx#1\next\def\tmp{#2}\widebarD 
   \else\expandafter\expandafter\expandafter\widebarC
   \fi\fi
}
\def\widebarD#1\end. {\fi\fi}

\def\widebarE{\widebarC A1.4 J1.2 L.6 O.8 T.5 U.7 V.3 W.1 Y.2 
   a.5 b.2 d1.1 h.5 i.5 k.5 l.3 m.4 n.4 o.6 p.4 r.5 t.4 v.7 w.7 x.8 y.8
   \alpha1 \beta1 \gamma.6 \delta.8 \epsilon.8 \varepsilon.8 \zeta.6 \eta.4
   \theta.8 \vartheta.8 \iota.5 \kappa.8 \lambda.5 \mu1 \nu.5 \xi.7 \pi.6
   \varpi.9 \rho1 \varrho1 \sigma.7 \varsigma.7 \tau.6 \upsilon.7 \phi1
   \varphi.6 \chi.7 \psi1 \omega.5 \cal1 \end. }

\def\tpbar{\widebar{\otimes}}

\def\PROOF{\noindent \quad{\csmc Proof \ }}
\def\RULE {\smallskip\noindent{\csmc Rule \ }}

\def\LEMMAnumbered#1{\smallskip\noindent {\csmc Lemma #1.}}

\def\moja{{1}}
\def\mbili{{2}}

\def\NatIsom{\cong}             



\phantom{}


   \PrintVersionNumber   


\sesquispace
\centerline{{\titlefont An inside view of the tensor product}\footnote{$^{^{\displaystyle\star}}$}%
{To appear in 
{\it\/Particles, Fields and Topology, Celebrating A.P. Balachandran\/},
 edited by T.R. Govindrajan, Guiseppe Marmo, V. Parameswaran Nair, Denjoe O'Connor, Sarada Rajeev and Sachindeo Vaidya
(World Scientific, Singapore)
}}


\bigskip


\singlespace			        

\author{Rafael D. Sorkin}
\address {Perimeter Institute, 31 Caroline Street North, Waterloo ON, N2L 2Y5 Canada}
\furtheraddress {Raman Research Institute, C.V. Raman Avenue, Sadashivanagar, Bangalore -- 560 080 India}
\furtheraddress {School of Theoretical Physics, Dublin Institute for Advanced Studies, 10 Burlington Road, Dublin 4, Ireland}
\furtheraddress {Department of Physics, Syracuse University, Syracuse, NY 13244-1130, U.S.A.}
\email{rsorkin@perimeterinstitute.ca}


\AbstractBegins                              
Given a vector-space 
$~V~$ 
which is the tensor
product of vector-spaces $A$ and $B$, we reconstruct $A$ and $B$ from the
family of simple tensors $a{\otimes}b$ within $V$.  
In an application to quantum mechanics, one would be reconstructing the
component subsystems of a composite system from its unentangled pure
states.
Our constructions can be viewed as instances of the category-theoretic
concepts of functor and natural isomorphism, and we use this to bring
out the intuition behind these concepts, and also to critique them.
Also presented are some suggestions for further work, including a
hoped-for application to entanglement entropy in quantum field theory.
\AbstractEnds

\bigskip


\let\parskipsave=\parskip
\parskip=1pt
\singlespace
\centerline {\bf Contents}

\item{1.} Introduction
\item{2.} {Posing the problem}
\item{3.} {The space $\S$ of simple vectors in $V$}
\item{4.} {How to recover $V_1$ and $V_2$ up to scale}
\item{5.} {The analysis and synthesis of a tensor product}
\itemitem{} {Using ``pointed vector spaces''} 

\vbox{
\item{6.} {Categorical matters (exposition and criticism)}
\itemitem{} {A small illustration: topology remembers spin}
\itemitem{} {A functorial gloss on our constructions}
}

\item{7.} {Questions; further developments; connection to quantum field theory}

\parskip=\parskipsave
\bigskip\medskip
\noindent
This article is dedicated to my friend, 
A.P. Balachandran, 
on the occasion of his 85th birthday.  
With his knack for discerning concrete implications of abstract
mathematical relationships, maybe he'll think of an unexpected use in
physics for the conception of tensor product proposed herein!

\bigskip
\bigskip


\sesquispace
\vskip -10pt

\section{1.~Introduction}                    
It would hardly be possible to review all the ways in which tensors
enter into physics.  General Relativity and Quantum Field Theory would
not exist without certain individual tensors or tensor-fields,
like the Lorentzian metric, the Riemann curvature, or the stress-energy
tensor, but it is perhaps in abstract quantum theory where the concept
of tensor-product itself, 
and of the corresponding {\it\/product-space\/} 
is most prominent.
The reason, of course, is that insofar as one deals with state-spaces of
quantal ``systems'', the tensor-product furnishes the construction that
combines the state-spaces of two or more subsystems into that of the
larger compound system or ``whole''.

Given this role of tensor-product, it could be unsettling that 
aside from its dimension, 
the resulting state-space (call it $V$) appears to remember
nothing about
the constituent spaces whose product it is.  For example,
because $12 = 4 \times 3 = 2 \times 6$,
a given state-space of dimension
12 that arose by combining spin $3/2$ with spin $1$
might equally well be describing a composite of spins $1/2$ and $5/2$.
Thus arises the following mathematical question.  

Suppose that a certain vector space $V$ is the tensor product of spaces
$V_1$ and $V_2$.  What extra information do you need in order to recover
$V_1$ and $V_2$ from $V$?  Or to put the question another way:
What does it mean for $V$ to carry the structure of a tensor product space?

To bring this question into sharper focus, imagine that instead of being
a tensor-product, $V$ were a direct sum, as it would be for example if
$V_1$ described an ionized Hydrogen atom, while $V_2$ described the same
atom in unionized form.  Then the appropriate state-space would be
$V=V_1\oplus{V_2}$; and the same equation would describe other mutually
exclusive alternatives, like an alpha particle being inside vs. outside
a nucleus, or a molecule being ortho-hydrogen vs. para-hydrogen.
To our question about tensor products, the counterpart in such cases would be: 
What does it mean for $V$ to carry the structure
of a direct sum?  Here however the answer is simple.  One only needs to 
indicate
%
$V_1$ and $V_2$ as {\it\/subspaces\/} of $V$. 
These two ``parts'' of $V$ are contained bodily within the whole, and every
$v\in V$ is uniquely a sum, $v=v_1+v_2$. 
A ``direct sum structure'' for $V$, 
in other words, 
is nothing but a pair of complementary
subspaces of $V$.  

Why isn't the case of a tensor-product space $V$ equally straightforward?
To appreciate what's different, let $V=V_1\otimes V_2$ be a tensor
product space.  The first thing one can notice is that $V_1$ and $V_2$
are no longer contained within $V$ in any obvious way.  Moreover, the
analog of $v=v_1+v_2$ in a direct sum, would be $v=v_1\otimes v_2$, but
how could one express this within $V$, given that vector-spaces are by
definition endowed with a notion of sum but not of product!
And lacking the operation $\otimes$ within $V$, how could you recover $V$
from $V_1$ and $V_2$, even if two such subspaces of $V$ could be identified?
%
%
The development that follows will answer these questions
and show that the space  $\S$ of {\it\/simple vectors\/} in $V$ 
--- elements of $V$ of the form $v=v_1\otimes v_2$ ---
can play the structural role that the complementary subspaces played in
the case of direct sum.

Although our questions are physically inspired, they are purely
mathematical, and the answers we will come upon must surely be well
known in some circles, even if they haven't shown up in the literature
I'm familiar with.
Nor do I know whether the constructions we will explore have any deeper
physical ramifications. They do, however, offer a more
intrinsic way to conceive of the tensor product 
(an ``inside view'' as one might say);
and a more intrinsic conception, more often than not, illuminates and deepens one's intuition.
%
%
As anyone who has taught a course in Relativity, Differential Geometry, or
``Mathematical Methods'' can attest, there's something about the concept
of tensor that intuition finds hard to grasp.  Perhaps it's no
coincidence then that quantum entanglement, the mathematics of which is
that of tensor product spaces, also seems counter-intuitive to so many
people.  

In this situation, the ``{\it\/square construction\/}'' on which our
development will rest, offers a complementary way to think about tensor
products, a way that starts not with the individual factor-spaces, 
$V_1$ and $V_2$,
but with the product space $V$ itself.  
Inasmuch as this more analytical approach shifts the main emphasis
onto the simple vectors within $V$, and inasmuch as these simple vectors
in a quantum context are precisely the unentangled state-vectors, our
development arguably makes better contact with physical intuition than the
more formal definitions one usually encounters.  At the very least, it
offers an alternative to more familiar ways of approaching the topic. 
After all, the more ways one has to think about a subject, the better
the prospect that at least one of them will be able to provide the key
to the deeper understanding that one is seeking.

The main ingredients of our constructions are presented in Sections 2-5
below.  Sections 2 and 3 are preparatory, while the constructions themselves
appear in Sections 4 and 5.  

In Section 6 we ask whether the results of Sections 4 and 5 can be
regarded as fully capturing the structure of $V$ as a tensor product;
and we show in some detail how to make this type of question precise in
terms of the category-theoretic concepts of functor and natural
transformation, to which we provide a brief introduction.  In this
connection, we also point out a certain shortcoming of the functor
concept itself.

In Section 7 we suggest some extensions of our constructions to tensor
products of three or more spaces, or to products of a single space with
itself, in which case symmetry-conditions come into play (bosonic,
fermionic, or nonabelian).  We also speculate that a suitable
generalization of the notion of simple vector to infinite dimensions
could help to clear away the mathematical obstructions
(involving type-III von Neumann algebras, where tensor products as
normally defined are not available)
that prevent one from understanding
entanglement-entropy (with its need for a cutoff) in terms of reduced
density matrices.





In what follows, we will assume,
unless otherwise specified,
that all vector spaces are real
and finite dimensional.
Nothing would change if we replaced the field $\Reals$ with the
field $\Complexes$, but it's convenient for exposition to pick one or the
other and stick with it.
We will also assume without special mention that the spaces $V_1$ and
$V_2$ are distinct from each other.

\section{2.~Posing the problem}              
Our question asks for a more intrinsic definition of tensor product,
or as we worded it above:
What could play the role of a tensor-product structure for $V$?
We will contemplate three possible answers to this question, 
%
%
and in doing so we will always assume that the two spaces of which $V$
is a product are distinct from each other.  Often this will not matter,
but sometimes it would make a difference, notably in definition (1) of
the paragraph after next, and then farther below in the ``Second
answer'' to our main question.

Before suggesting answers to our main question, however, it seems
advisable to dwell for a moment on the more common definitions of tensor
product.
Different authors favor different ones, and 
the answers to our questions will tend to take on different forms,
depending on which definition one has in view.  What, then,
are some of the popular definitions 
[1]
of the space $V_1{\otimes}V_2\,$ and of the tensors therein?

\noindent (1) an element of $V_1{\otimes}V_2\,$ is a numerical matrix
whose entries depend on a choice of bases for $V_1$ and $V_2$ 
and
transform in a certain way when these bases are changed.  (This might be
the oldest definition.)

\noindent
(2) an element of $V_1{\otimes}V_2\,$ is a linear mapping between two
 vector spaces, for example a linear mapping from $V_2^*$ to $V_1$,
 where $V_2^*$ is the dual space of $V_2\,$

\noindent(3) an element of $V_1{\otimes}V_2\,$ is an equivalence class
 of formal sums of symbols, 
$\alfa{\otimes}\beta\,$, where $\alfa\in V_1$ and $\beta\in V_2$

\noindent (4) ``the'' space $V_1{\otimes}V_2$ is any solution of a certain
 ``universal mapping problem'' involving bilinear functions from
 $V_1\times V_2$ to an arbitrary vector space $Y$.  (Thereby a
 bilinear function from $V_1 \times V_2$ to $Y$ induces a unique linear
 mapping from $V_1{\otimes}V_2$ to $Y$.)

\smallskip\REMARK  Notice that definition (1) refers to independently chosen bases 
for $V_1$ and $V_2$.  Were we to assume that $V_1$ and $V_2$ were
literally the same space, only a single basis would come into play.
Tensors in this vein are common in GR and differential geometry, with
$V_1=V_2$ being the tangent space to a point of spacetime.  In quantum
mechanics on the other hand, distinct factor-spaces are typical, albeit
not in the case of indistinguishable particles.

In finite dimensions, all these definitions are provably equivalent.
In infinite dimensions one must distinguish between the so-called
 algebraic tensor product and various topological tensor products, not
 all of which are the same in general.
For our purposes, entering into those subtleties would be too much of a
distraction.  [2]
Instead, we will work throughout in finite dimensions.
In the jargon of category theory, each of
these definitions yields a ``functor'' from pairs of vector spaces to
vector spaces, and the statement that 
(in finite dimensions) 
they are all equivalent 
asserts that between any two of the functors 
there is an invertible ``natural transformation''. 
Now back to our main question and some possible answers to it.

\noindent {\it\/First possible answer\/}.
The most direct and obvious answer, but at the same time the least
informative, is that a tensor-product structure for $V$ is an
isomorphism between $V$ and a space of the form $V_1{\otimes}V_2$.  
This is a good start, but it has the drawback that the auxiliary
spaces $V_1$ and $V_2$ are not derived from $V$, with the consequence
that different choices of them would strictly speaking define different
tensor-product structures for $V$.  We could address this difficulty by
forming equivalence-classes under isomorphisms of $V_1$ and $V_2$, but
let's instead continue on to the second and third proposals.

\noindent {\it\/Second answer\/}.
The second possible answer to our main question, already more concrete
and ``intrinsic'', hearks
back to definition (1) in the above list. 
In order to represent an
element of $V=V_1{\otimes}V_2$ as a numerical matrix, one needs a basis of $V$
whose
members are themselves organized into a matrix.  Specifically, if a list of
vectors $e_j\in V_1$ furnishes a basis for $V_1$ and a second list of 
vectors $f_k\in V_2$ furnishes a basis for $V_2$, then the products
$e_j{\otimes}f_k$ furnish a basis for $V_1{\otimes}V_2$ whose members
array themselves in a rectangular matrix with rows labeled by $j$ and columns by $k$.
In such a basis the matrix representing a simple vector
$v_1{\otimes}v_2$ will be a matrix-product of the form,
column-vector $\times$ row-vector.
(Such a ``special basis'' is precisely an isomorphism between $V$ and
$\Reals^m {\otimes} \Reals^n$, where $m$ and $n$ are the respective
dimensions of $V_1$ and $V_2$.)

We could thus answer that a tensor-product structure for $V$ is a basis
for $V$ organized into a rectangular matrix.  Unfortunately this won't
quite do, because many other bases will define the same product-structure.
First of all, 
one might swap rows with columns, which 
amounts to writing $V_2{\otimes}V_1$ instead of $V_1{\otimes}V_2$.  This
does nothing.  
However one can also replace each of the two bases by some other basis 
for the same space, 
which doesn't change the spaces $V_1$ or $V_2$ themselves but only their
representations.  We are thus led to identify a tensor-product structure
for $V$ as an equivalence-class of bases, parameterized by
$G=GL(n_1)\times GL(n_2)$ where $n_i=\dim(V_i)$.  But even this is not
quite correct since it is always possible to rescale the basis for $V_1$
by some factor, while rescaling the basis for $V_2$ in the opposite way.
Since this doesn't affect the resulting basis for $V_1{\otimes}V_2$, we
conclude that $G$ is really the quotient group
$GL(n_1){\times}GL(n_2)\,/\,GL(1)$.  
Here of course $GL(n)$ is the group of invertible $n\times n$ matrices.

The fact that true group is $GL(n_1)\times GL(n_2)\,/\,GL(1)$ and not simply
$GL(n_1)\times GL(n_2)$ seems a detail, but it is actually telling us
something that will show up again in our deliberations below.  From a
tensor-product structure for $V$, we cannot fully reconstruct the
factor-spaces $V_1$ and $V_2$; we can obtain them only up to a joint
scaling ambiguity.

\REMARK In a classical (non-quantum) context, a composite system would be
described by a cartesian product, $A\times B$.  In that case, the
counterpart of a rectangular basis would just be 
(for discrete spaces $A$ and $B$) 
a rectangular list of elements of $A\times B$, 
and the story would more or less end there.\footnote{$^\star$}
{The counterpart of $G$ would be the product of the permutation groups of $A$ and $B$}
Tensor products are more subtle than cartesian products, however,
and there's a third possible answer to our question which is 
still more concrete and 
intrinsic than an equivalence class of bases.

%
%

\noindent {\it\/Third answer\/}.
The third possible answer to our question, 
and the one which the rest of this paper will explore, 
is that the tensor-product structure for  $V=V_1{\otimes}V_2$ can be
taken to be the subspace $\S$ of {\it\/simple vectors\/}:
$$
        \S = \SetOf{\alfa \otimes \beta}{\alfa \in V_1, \beta \in V_2}  \eqno(1)
$$
As we will see, there exist explicit constructions that take you from 
$\S\subseteq V$ to 
(copies of)
$V_1$ and $V_2$, 
and thence back to $V$.

\REMARK In quantum language,  $\S$ would be the set of unentangled
state-vectors.  Obviously there's something special about them, but
mathematically they are only the first in a hierarchy of successively
more generic tensors, those of ranks 2, 3, etc, where the ``rank'' of
$v$ is the minimum number of simple vectors of which it is a sum
(quantum mechanically the number of terms in a Schmidt decomposition of
$v$.)\footnote{$^\dagger$}
{Algebraic Geometry has given the name ``Segre variety'',  not quite to $\S$ itself, but to the set of rays in $\S$}

\section{3.~The space $\S$ of simple vectors in $V$}         
Since it is related quadratically to $V_1$ and $V_2$, the set $\S$ of
simple vectors obviously will not be a linear subspace of $V$ in
general, but it will be foliated by two families of linear subspaces,
which we will denote by $\M_1$ and $\M_2$.

Before demonstrating this, let us deal with two trivial cases that don't
fit easily into the general pattern, 
In the most trivial case, both $V_1$ and $V_2$ are one-dimensional:
$\dim(V_1)=\dim(V_2)=1$.  Both are then copies of $\Reals$, as also is
$V=V_1 \otimes V_2$.  In this case every $v\in V$ is plainly a simple
vector, and so $\S$ is all of $V$.  
Conversely, given that $\dim(V)=1$, and since we know in general that
$\dim V_1 \otimes V_2 = \dim V_1 \times \dim V_2$, we know immediately
that both $V_1$ and $V_2$ are isomorphic to $V$ itself.  In a
reconstruction of $V_1$ and $V_2$, we can thus do no better than to take
both to be copies of $V$, and this suffices.  
The only small subtlety shows up when, 
having passed from $V$ to $V_1$ and $V_2$,
we seek to reconstruct $V$ as $V_1 \otimes V_2 = V \otimes V$,
but have to face the fact that 
although $V$ is isomorphic to $V \otimes V$, 
the isomorphism is not canonical.\footnote{$^\flat$}
{In some sense this is just ``dimensional analysis''.  If the elements
 of $V$ were ``lengths'', then those of $V \otimes V$ would be squared
 lengths.}
%




In the second trivial case, $\dim V_1>1$ while $\dim V_2=1$ (or vice versa).
Here again $\S$ is trivially all of $V$. 
(By definition 
any $v\in V$ is a sum of terms of the form $a\otimes b$ for 
$a\in V_1$ and $b\in V_2$, but since all nonzero $b$ are proportional to
each other, all the $b$ can be taken equal, whence
$v=a_1\otimes b + a_2\otimes b + \cdots = (a_1+a_2\cdots)\otimes b\in\S$.)
It follows that
$V=V_1 \otimes V_2 \isom V_1 \otimes \Reals = V_1$ 
(where `$\isom$' signifies isomorphic-to).  
Conversely, whenever $\S=V$, we can construct spaces $V_1$ and $V_2$
by taking $V_1$ to be $V$ and $V_2$ to be any one-dimensional vector space,
for example the subspace of $V$ given by $\Reals\,b_0$, where $b_0$ 
is any fixed vector\footnote{$^\star$}
{The reason for this particular choice will become clear soon.  Notice
 also that we could of course have exchanged the roles of $V_1$ and $V_2$.}
in $V$. 
The now-familiar scaling-ambiguity corresponds then to the undetermined
normalization of $b_0$.

Notice in these two examples that $V_1$ was identified with a maximal
linear subspace of $\S$.  Although completely trivial in the two
examples, this observation will be the basis of our reconstruction of
$V_1$ and $V_2$ in the generic case.
In seeking to understand the linear subspaces of $\S$, we will need a
few ``obvious'' facts about tensor products which we will now review in
the spirit of definition (3) mentioned in the previous section.

Recall then that any tensor $T \in V_1 {\otimes} V_2$ is a sum of
simple tensors, i.e. a sum of products of vectors from $V_1$ with
vectors from $V_2$:
$$
     T \,=\, \sum_j  \, a_j \otimes\, b_j      \eqno(2)
$$
To fully characterize the space $V_1 {\otimes} V_2$, however,
one needs to specify which such sums are equal to which others, 
or equivalently which expressions $\,T\,$ equal the zero tensor.
Intuitively the answer is that $T=0$ iff it is forced to vanish by the
combining rules for the symbols $a {\otimes} b$ together with the linear
dependences among the vectors of $V_1$ and $V_2$.  This criterion is
implicit in the aforementioned definition (4), but it is more useful to
express it algorithmically.

\RULE  Provided that the vectors $a_j$ in (2) are
linearly independent, $T=0$ if and only if all of the $b_j$ vanish.

\noindent (Obviously the same rule will hold true if we exchange the roles of $a_j$ and $b_j$.)
As stated, the rule wants the $a_j$ to be linearly independent.  If
they are not, then some of them can be expressed as linear combinations
of the others, and one should do this before applying the rule.  Thus,
an {\it\/algorithm\/} for deciding whether $T=0$ consists in first writing
any redundant $a_j$ in terms of the others, second expanding out the resulting
expression to put $T$ into the form (2),
and third applying the rule as stated.

As a trivial consequence of this rule, we learn that $a {\otimes} b$ is
nonzero if both $a$ and $b$ are.  In stating the following further
consequences, we will interpret $a\propto b$ to mean that either
$a=\lambda b$ or $b=\lambda a$, $\lambda\in\Reals$.

\LEMMAnumbered{\moja}
If $\alfa$ and $\beta$ are nonzero then 
$\alfa {\otimes} \beta = \alfa' {\otimes} \beta'$ 
$\implies$
$\alfa' \propto \alfa$ {\it\/and\/} $\beta' \propto \beta$.

\PROOF  Were $\alfa'$ not proportional to $\alfa$, they would be
linearly independent.  Our ``Rule''  would then imply that
$\alfa {\otimes} \beta - \alfa' {\otimes} \beta'$
could not vanish.  Therefore $\alfa' \propto \alfa$, and by symmetry
$\beta' \propto \beta$.

\LEMMAnumbered{\mbili}
  Let $\alfa \otimes \beta\in\S$ and $\alfa' \otimes \beta'\in\S$
be nonzero simple vectors.
If their sum is also simple then 
{\it\/either\/} $\alfa' \propto \alfa$ {\it\/or\/} $\beta' \propto \beta$.


\PROOF (by contradiction).  
Assume that $\alfa'\not\propto\alfa$ and $\beta'\not\propto\beta$.
The four terms $\alfa{\otimes}\beta$, $\alfa'{\otimes}\beta'$,
$\alfa{\otimes}\beta'$, $\alfa'{\otimes}\beta$
are then  (by a simple application of the Rule)
linearly independent.
By hypothesis
$\alfa\otimes\beta+\alfa'\otimes\beta'=\gamma {\otimes} \delta$
for some $\gamma$ and $\delta$.  Appealing once again to the Rule,
and remembering that $\alfa'$ is independent of $\alfa$, we conclude that 
$\gamma$ must be a linear combination of $\alfa$ and $\alfa'$; 
similarly 
$\delta$ must be a linear combination of $\beta$ and $\beta'$.
But then $\gamma {\otimes}\delta$,
when expanded out,
could not contain the required terms, 
$\alfa\otimes\beta$ and $\alfa'\otimes\beta'$ without also containing
terms in  $\alfa\otimes\beta'$ and $\alfa'\otimes\beta$.


Returning now to the analysis of $\S$,
and recalling that we have already disposed of the possibility that
either $\dim{V_1}=1$ or $\dim V_2=1$,
we can assume for now that
 $\dim{V_1}\ge2$ and $\dim V_2\ge2$, 
this being where the typical structure of $\S$ reveals itself,
namely that of the two foliations $\M_1$ and $\M_2$ already alluded to
but not yet defined.  
For the time being, we will define $\M_1$ and $\M_2$ as follows.
Soon,
we will define them intrinsically
(meaning directly from $V$ and $\S$ alone), 
whereupon
(3) and (4)
will shed their status as definitions and become theorems.
The members of $\M_1$ will be the subsets of $\S$ of the
form $V_1{\otimes}\beta$ for $\beta\in V_2$, and likewise for $\M_2$:
$$
       \M_1 = \SetOf{V_1 {\otimes}\,\beta}{\beta\in V_2}  \eqno(3)
$$
$$
       \M_2 = \SetOf{\alfa \, {\otimes} V_2}{\alfa \in V_1} \eqno(4)
$$
(Here of course our notation means that, e.g, $V_1 {\otimes}\,\beta = \SetOf{\alfa {\otimes}\,\beta}{\alfa \in V_1}$.)

We want to prove 
first, that every $M$ in either $\M_1$ or $\M_2$ is a {\it\/maximal linear subspace of\/} $\S$;
second that $\M_1$ and $\M_2$ exhaust the maximal linear subspaces of $\S$;
third that $M,N\in\M_1$ and $M\not= N$ $\implies M\cap N = \braces{0}$ 
(and likewise for $\M_2$);
and 
fourth that $M\in\M_1, N\in\M_2 \implies \dim( M\cap N)=1$.


Why are the members of $\M_1$ (for example)  maximal linear subspaces of $\S$?  
That $M=V_1 {\otimes}\,\beta$ is a linear subspace is obvious, but why is it maximal?  
Well, any simple vector not in $M$  must take the form $\alfa'{\otimes}\beta'$ 
with $\beta'$ independent of $\beta$.  
Choose also an $\alfa\in V_1$ that is
independent of $\alfa'$ (which is always possible since $\dim{V_1}>1$),
and notice that $\alfa{\otimes}\beta\in{M}$.
If we could adjoin $\alfa'{\otimes}\beta'$ to $M$ then 
$\alfa'{\otimes}\beta'+\alfa{\otimes}\beta$ would also have to be
in $M$, and therefore simple, contrary to Lemma {\mbili} above.\footnote{$^\dagger$}
{What we are effectively proving could be reduced to a lemma to the effect that
 every linear subspace of $\S$ has the form $W{\otimes}\beta$ or
 $\alfa{\otimes}W$, for some vector-subspace $W$ of $V_1$ or $V_2$}
%

And why does every maximal linear subspace of $\S$ have to belong to
either $\M_1$ or $\M_2$?  Well, let $M$ be such a subspace, and let
$\alfa {\otimes} \beta\in M$.  Certainly $\alfa {\otimes} \beta$ alone
is not maximal 
(it belongs to 
$V_1 {\otimes}\,\beta$, for example),
so let $\alfa'{\otimes}\beta'$ be an independent
member of $M$.  By the same lemma 
either $\alfa$ and $\alfa'$ are proportional 
or $\beta$ and $\beta'$ are proportional, 
say the latter.  Then as
we just saw, every other member of $M$ must also take the form 
$\gamma {\otimes} \beta$ for some $\gamma\in V_1$, in other words 
$M\subseteq V_1 {\otimes}\,\beta\in \M_1$, whence 
$M=V_1 {\otimes}\,\beta$ since $M$ is maximal.

Third, if $M,N\in\M_1$ are unequal then $M=V_1 {\otimes} \beta$ and
$N=V_1 {\otimes} \beta'$ with $\beta$ independent of $\beta'$.  
Hence any $v\in M\cap N$ must satisfy 
$v=\alfa {\otimes} \beta = \alfa' {\otimes} \beta'$
for some $\alfa$ and $\alfa'$. 
But by Lemma \moja, this is impossible unless $v=0$.

Fourth, if $M\in\M_1, N\in\M_2$ then $M=V_1 {\otimes} \beta$,
$N=\alfa {\otimes} V_2$ for some $\alfa\in V_1, \beta\in V_2$.  
If $v\in M\cap N$ then by definition, 
$v=\alfa' {\otimes} \beta = \alfa {\otimes} \beta'$
for some  $\alfa'\in V_1, \beta'\in V_2$.
The lemma just cited then informs us that 
$\alfa'\propto\alfa$ and $\beta'\propto\beta$, whence 
$v=\alfa' {\otimes} \beta \propto \alfa {\otimes} \beta$.
In other words $M\cap N$ is the 1-dimensional subspace, 
$\,\Reals \, \alfa {\otimes} \beta$

The essential feature  we have discovered is that 
{\it\/any two members of different foliations meet in a ray (a one-dimensional subspace of $V$) and any two distinct members of the same foliation are disjoint.\/}
This lets us determine the foliations $\M_1$ and $\M_2$
simply from a knowledge of $\S\subseteq V$, 
without any further recourse to how
$V$ arose as a tensor product: if $M$ and $N$ are elements of the set
$\M$ of all maximal linear subspaces of $\S$,
then they belong to the same foliation if and only if they are disjoint,
and this criterion is guaranteed to produce exactly two disjoint subsets
of $\M$, which we can label as $\M_1$ and $\M_2$.
Henceforth, we will adopt this
{\it\/intrinsic definition\/} of $\M_1$ and $\M_2$, 
which we can record in the following two maps that associate
with each simple vector in $V$ the two maximal linear subspaces of $\S$
to which it belongs.  

\DEFINITION  Let $v\in\S$.  Then $\pi_1(v)$  [resp. $\pi_2(v)$] is the
unique maximal linear subspace of type $\M_1$ [resp. $\M_2$] that contains $v$.  

\noindent
Equations, (3)-(4), are hereby no longer definitions but
theorems which apply 
whenever we can exhibit
vector-spaces $V_1$ and $V_2$ such that $V=V_1{\otimes}V_2$.

With these observations, we have taken
a first step in recovering the tensor
product structure of  $V$ from $\S$.  In fact, one sees from (3) and
(4) that 
each $M_1$ in $\M_1$ is a copy of $V_1$
and
each $M_2$ in $\M_2$ is a copy of $V_2$.
In the following section, we will build on
our knowledge of $\M_1$ and $\M_2$
to recover fully the ray-spaces
associated to $V_1$ and $V_2$,
and then to recover $V_1$ and $V_2$ themselves up to scale.

\section{4.~How to recover $V_1$ and $V_2$ up to scale}      
Our ultimate aim is to find a construction that, 
relying on nothing more than the set $\S$ of simple vectors in $V$, 
will resolve the latter into its two factors (as uniquely as possible),
and then to discover how to rebuild  $V$ as the tensor product of these factors.
%
%
This will take place in Section 5, and not everything from the
present section will be needed there.  If you are reading these lines,
you might thus want to skip over the present section in order to appreciate the
great simplicity of the final constructions.  On the other hand, the
present section, as well as providing much of the background for Section 5, will 
also show how, in becoming aware of the two spaces $\M_1$ and $\M_2$, 
we have {\it\/already\/} recovered from $\S$ the {\it\/rays\/} of
$V_1$ and $V_2$, which in a quantum context means we have already recovered,
if not the respective subsystems themselves, 
then at least their ``pure states''.

To appreciate this fact, recall that when $V=V_1 {\otimes} V_2$, any member
$M$ of $\M_2$ can be expressed in the form (4).  But the subspace 
$M=\alfa\,{\otimes}V_2$ determines and is determined by the ray,
$\Reals\,\alfa\subseteq V_1$.  The points of $\M_2$ are thus in bijective
correspondence with the rays of $V_1$, and likewise for $\M_1$ and $V_2$.
%
%
Introducing the notation $\P{V}$ for the projective space
formed from the rays of any vector-space  $V$,
we can therefore assert 
that 
$$
    \P V_1 = \M_2 \qquad \hbox{and} \qquad \P V_2 = \M_1  \eqno(5)
$$
Of course, there's more to it than this, because so far, we have only
introduced $\M_1$ and $\M_2$ as sets without further structure.  
In order to fully corroborate the claim that $\P V_1 = \M_2$, we need to present $\M_2$ 
as the set of rays of some intrinsically defined  vector space,
this being one way to equip it with a projective structure.
In the course of doing so, we will also see 
how to get our hands on $V_1$ itself up to scale. 

Let's first see the procedure per se and then return to see more fully why it works.
To get started, select arbitrarily 
any $M\in\M_1$
and let $P$ be the restriction of $\pi_2$ to $M$.
It is not hard to see that 
$P:M\to\M_2$
sets up a one-to-one correspondence between the rays in $M$ and the points of $\M_2$. 
%
By definition, if $v\in M$ then $P(v)=\pi_2(v)$ is the unique maximal linear
subspace in $\M_2$ that contains $v$; being linear, it also contains the entire
ray, $\Reals\, v$.  Furthermore, $P$ is trivially surjective because for
any $N\in\M_2$,  $M\cap N$ is (as observed earlier) a ray $\ell$ in $M$ that
gets mapped by $P$ to $N$ itself.  This also proves that $P$ is
injective (on the {\it\/rays\/} of $M$)  because 
any other ray in $M$ that was mapped to $N$ by $P$ would by
definition have to lie in $M\cap N$ and therefore coincide with $\ell$.

The mapping, $P:M\to\M_2$, is what we were looking for, 
but it remains to demonstrate that any other $M'\in\M_1$
would have induced the same projective structure on $\M_2$.  
For this, it suffices to find a linear isomorphism between 
$M$ and $M'$
that commutes with the corresponding projections.
In other words,
with $P'$ taken to be the restriction of $\pi_2$ to $M'$,
we should seek an isomorphism $f:M\to M'$ such that $P=P'\compose f$.
Such an $f$ would induce an isomorphism $\fhat: \P M\to\P M'$, and it is
actually easier to characterize this isomorphism intrinsically than to exhibit $f$
itself.  Let us therefore define $\fhat$ first, and only then consider
how to lift it to a linear map $f$.
It turns out that the Ansatz,
$$
    \fhat(M \cap  N) = M' \cap N  \eqno(6)
$$
%
(where  $N$  is an arbitrary element of $\M_2$)
does what is needed.  
In particular, 
if $M''$ is a third element of $\M_1$, 
then the
isomorphisms $M \to M' \to M''$ 
defined by (6) obviously compose consistently.

Our remaining task is to lift the just-constructed mapping, 
$\fhat: \P M \to \P M'$,
to a linear function,
$f : M\to M'$.
Given that for $v\in M$, 
the mapping $\fhat$ already determines
the ray in $M'$ to which $v$  should go, the only further input needed 
to define $f(v)$
is its
normalization.  
Although this seems a tiny bit of extra
information, the construction 
via which we will obtain it
is surprisingly intricate.  
In fact, it is not really needed for present purposes;
all we really need to know is that a linear lift $f$ exists, 
which could be proven more easily.  
If nevertheless we take
the trouble to construct $f$ explicitly, it is because 
doing so
will introduce us to a 
certain type of 
``simple-square''
that will play an important role in the
next section.

Fix spaces $M,M'\in\M_1$ as above, and let 
$\ell_0$ be any ray in $M$, 
with $\ell_0'=\fhat(\ell_0)$ being the corresponding ray in $M'$,
as given by (6).
We know that $f$ will take any point in $\ell_0$ to some point in $\ell_0'$.
Given now some arbitrarily chosen reference vector,
$v_0\in\ell_0$,  
we need to 
decide which vector in $\ell_0'$ will be $f(v_0)$,
and it turns out that this decision determines $f$ fully.
Let  $v_0'$ be the vector selected to be $f(v_0)$. 
The problem then is to determine $f(v)$
when $v$ belongs to some other ray $\ell\subseteq M$.  
That is, we need to figure out where $f(v)$
lies along the ray $\ell'=\fhat(\ell)$.

This problem admits a generic case and a couple of special cases.
In the generic case, $v_0$, $v_0'$, and $v$ are all linearly
independent.  Consider then 
an arbitrary  $v'\in \ell'$ and
the {\it\/square\/}
$$
         \pmatrix{a & b \cr
                    &   \cr
                  c & d \cr}            
   = 
         \pmatrix{v_0 & v_0' \cr
                      &      \cr
                  v   & v'   \cr}
\eqno(7)
$$
whose elements belong to the rays
$$
          \pmatrix{\ell_0 & \ell_0' \cr
                          &   \cr
                     \ell & \ell' \cr} \ .  \eqno(8)
$$
By construction (cf. (6)), 
$$
 \pi_1a=\pi_1c \,, \quad 
 \pi_1b=\pi_1d \,, \quad 
 \pi_2a=\pi_2b \,, \quad 
 \pi_2c=\pi_2d \,.        \eqno(9)
$$
Consequently, the two row-sums and the two column-sums belong to $\S$
(i.e. all four sums are simple vectors in $V$), 
but what about the overall sum, $a+b+c+d\,$?
In the answer to this question lies the key to our construction of $f$.
In fact 
(as we will prove shortly)
this sum meets $\S$ for precisely one point $v'$ in the ray
$\ell'$, and by setting $f(v)=v'$ we define $f$ uniquely on the ray
$\ell$.
%
Doing the same for the other rays in $M$, we will obtain a
function $f:M\to M'$ which is linear,
unique up to a multiplicative prefactor, and whose action on rays is 
by definition that of $\fhat$.

So much for the generic case.
Before turning to the special cases, observe that just from (9)
alone, we can write the rays in (8) as
$$
          \pmatrix{M \cap N  &  M' \cap N \cr
                             &            \cr
                  M \cap N'  &  M' \cap N' \cr} \ ,  \eqno(10)
$$
where 
 $M =\pi_1a=\pi_1c$,
 $N =\pi_2a=\pi_2b$,
 $M'=\pi_1b=\pi_1d$,
 $N'=\pi_2c=\pi_2d$.
%
%
%
%
The generic case just treated corresponded to an array
(10) in which the subspaces, 
$\,M,M',N,N'\,$, 
were all distinct, and correspondingly the vectors, 
$\,a,b,c,d\,$, in (7) were linearly independent.
The special cases we still need to treat are those in which $M=M'$ or $N=N'$.

Consider first the special case 
where $N=N'$,
or equivalently, 
$\ell=\ell_0$. 
%
Here
we know the answer
trivially because $v=\lambda v_0$ for some scalar $\lambda$, 
whence 
$v'=f(v)=f(\lambda v_0)=\lambda f(v_0) = \lambda v_0'$.
The square in (7) thus assumes the form,
$$
         \pmatrix{a        & b \cr
                           &   \cr
                 \lambda a & \lambda b \cr}    \eqno(11)
$$
The other special case is that 
where $M=M'$, 
or equivalently (since, as we know, any two elements of $\M_1$ are either equal or disjoint),
$\ell_0'=\ell_0$.
Here $f$ is just mapping $M$ to itself, an obvious solution for which
would be to  take $f$ to be the identity map.  However, we could equally well
take it to be a multiple of the identity by a scalar $\mu$, in which case our
square would take on the appearance,
$$
         \pmatrix{a        & \mu a \cr
                           &   \cr
                 c & \mu c \cr}  \ ,    \eqno(12)
$$
a form that follows immediately from (11) by symmetry. 
For completeness, let us also record the doubly special case where
$M=M'$ and $N=N'$ 
both hold, leading to a square of the design,
$$
         \pmatrix{a        & \mu a \cr
                           &   \cr
                  \lambda  a & \lambda \mu a \cr}  \ ,    \eqno(13)
$$
as one sees by combining (11) with (12).
All these special cases, (11) --
(13), can be obtained from the generic case
by forming limits.
%
%
Amalgamating these special cases with the generic one,
we arrive at the following definition.

\medskip\DEFINITION
A {\it\/square\/} (or {\it\/simple square\/}) 
is a matrix $\pmatrix{a&b\cr c&d}$ of simple vectors 
which satisfy (9), 
and which 
in the generic case satisfy $a+b+c+d\in\S$, 
or in the special cases take on one
of the forms (11)--(13). 
%


\smallskip\noindent
The reason for separating the generic from special cases in the definition is that 
$a+b+c+d\in\S$ suffices in the generic case, but not in the special cases.
Of course, it holds in the latter cases too, albeit it is trivial there.
It's also worth noting that given any simple square, 
one can multiply any row or
column by a scalar without invalidating its status as a square.
And for completeness, let us recall from above that the two row-sums and
the two column-sums also belong to $\S$.

This completes the description of our procedure for defining $f$.
In order to understand why it works, 
let's ``look behind the curtain'' to see what our squares
amount to when
expressed in terms of vectors in $V_1{\otimes}V_2$.
(This should also help to illuminate the rather abstract development we have been following in this section.)
Recall that the four rays in (8) can also be written as
the intersecting subspaces exhibited in (10).
Now by equations (3) and (4), 
$M=V_1{\otimes}\beta_0$ for some $\beta_0\in V_2$, while  
$N=\alfa_0{\otimes}V_2$ for some $\alfa_0\in V_1$,
and similarly 
$M'=V_1{\otimes}\beta$,
$N'=\alfa{\otimes}V_2$,
for some $\alfa$ and $\beta$.  
Without loss of generality we can therefore write (7) in the
form,
$$
         \pmatrix{a & b \cr
                     &   \cr
                  c & d \cr}            
   =
         \pmatrix{\alfa_0 \,\otimes\,\beta_0  & \alfa_0 \,\otimes\, \beta  \cr
                     &   \cr
                  \alfa \,\otimes\,\beta_0 &  \lambda \,\alfa\,\otimes\,\beta\cr}
\eqno(14)
$$
where {$\lambda$} is some unknown coefficient of proportionality.
This form makes it plain that the row- and column-sums are  indeed
simple, for example  $a+b=\alfa_0 {\otimes}(\beta_0+\beta)$.
As for the overall sum, 
$a+b+c+d$,
it will be the simple vector,
$(\alfa_0+\alfa){\otimes}(\beta_0+\beta)$ 
{\it\/provided that\/} $\lambda=1$.
Were $\lambda\not=1$ on the other hand,
the same sum would equal
$(\alfa_0+\alfa){\otimes}(\beta_0+\beta)+(\lambda-1)\alfa {\otimes}\beta$,
which according to Lemma \mbili, could be simple only if 
$\alfa$ were proportional to $\alfa_0$ or 
$\beta$ were proportional to $\beta_0$, meaning we'd be back in one of
the special cases we disposed of earlier.

In summary, consider a square of simple vectors belonging to rays of the
form exhibited in (10)
with $M,M'\in\M_1$ and $N,N'\in\M_2$,
and assume we are in the
generic case where $M\not=M'$, $N\not=N'$.  On condition that the sum of
all four simple vectors is itself simple, any three of them determine
the fourth uniquely.
The vectors must in that case ``secretly'' take the form
(14) with $\lambda=1$:
$$
         \pmatrix{\alfa_0 \,\otimes\,\beta_0  & \alfa_0 \,\otimes\, \beta  \cr
                     &   \cr
                  \alfa \,\otimes\,\beta_0 &  \alfa\,\otimes\,\beta\cr}
\eqno(15)
$$
Our special cases correspond 
to $\alfa_0\propto\alfa$ and/or $\beta_0\propto\beta$,
and they also fit the pattern (15), 
which accordingly represents the universal form 
that a square assumes 
when one views it ``from behind the curtain''.

Returning 
to the task of lifting
$\fhat: \P M\to\P M'$  
to a linear  isomorphism, $f:M \to M'$,
we can now see that 
the construction of $f(v)$
following equation (9)
does indeed do the job,
because it maps
$M=V_1 {\otimes} \beta_0$ to $M'=V_1 {\otimes} \beta$
by carrying
$\alfa{\otimes}\beta_0\in M$ to $\alfa{\otimes}\beta\in M'$, 
a correspondence which is plainly linear
when $\alfa$ varies.
Of course, the fact that $f$ is a lift of $\fhat$
cannot determine it uniquely,
because any multiple of a lift is another lift.
It's thus no accident that our 
construction
involved a free choice of reference vectors,
$v_0$ and $v_0'$.  
A different choice, however, could only alter $f$ by an overall factor,
as follows from the general fact that any two linear isomorphisms that
induce the same mapping on rays must agree up to scale.\footnote{$^\flat$}
{Proof. Call the maps $f$ and $g$ and let $x$ and $y$ be any two
 independent vectors in their domain with $z=x+y$.  By assumption
 $g(x)=\lambda f(x)$ and $g(y)=\mu f(y)$, and we want to prove that
 $\mu=\lambda$.  By rescaling either $f$ or $g$ if necessary, we can
 assume that $\lambda=1$.  But then $f(z)=f(x)+f(y)$ would lie in a
 different ray from $g(z)=f(x)+\mu f(y)$ unless $\mu=1$ as well. }
For the same reason, we don't need to check our isomorphisms $f$ for  
coherence.
Given that they cohere on $\P M \to \P M' \to \P M''$, 
as we already know they do,
they must also cohere up to scale on $M \to M' \to M''$,
and that's the best we can do.


Taking as input solely the set $\S$ of simple vectors in $V$, we have now
identified with one another the members of $\M_1$ via isomorphisms which
are unique up to scale.  On one hand we used this to 
derive from $V$-cum-$\S$ a canonically given projective space
that is
naturally isomorphic to $\P V_1$ (the space of ``pure states of system-1'' in a quantal interpretation).
On the other hand, these same identifications produce
a vector
space that is naturally isomorphic to $V_1$ itself, albeit only modulo a
scaling ambiguity.  The same procedure applied to $\M_2$ rather than
$\M_1$ would obviously recover $\P V_2$ and $V_2$ in the same sense.
Our task now is to complete the story by re-building $V$ as the tensor
product of the two vector spaces just constructed.

\section{5.~The analysis and synthesis of a tensor product}  
%
%
%
Our previous work has already led us to pay close attention to the maximal
linear subspaces of $\S \subseteq V = V_1\tensor V_2$. 
%
%
Let us now select two such spaces,
$W_1\in\M_1$ and $W_2\in\M_2$,
and then select further a vector $w_0\in W_1\cap W_2$ to serve as their common
``base point''.\footnote{$^\star$} 
{In Sections 3 and 4 we usually used the letters $M$ and $N$ to denote maximal linear subspaces of $\S$.
 The notation, $W_1$, $W_2$,  here is chosen to emphasize the parallelism with $V_1$, $V_2$.}
We have then
$$
   W_1=\pi_1 w_0 \ ,   \qquad W_2=\pi_2 w_0 \ .
$$
We want to demonstrate
that $V$  can be 
construed
as the tensor-product
of these two spaces.
%
%

To that end, 
and basing ourselves on
the  concept of ``square'' introduced in Section$\,4$,
we will 
introduce a new bilinear product, $\tpbar: W_1\times W_2\to V$,
as follows.
For $w_i\in W_i\ (i=1,2)$, 
let us
define 
$\,w_1\tpbar w_2\,$ 
to be the solution of
the following square:
$$
         \pmatrix{w_0 & w_2 \cr
                     &   \cr
                  w_1 & w_1\tpbar\, w_2 \cr}   \eqno(16)
$$
In other words,
%
$w=w_1\tpbar \, w_2$ 
must satisfy
the conditions,
$$
   \,w\in \pi_2 w_1 \,\cap\, \pi_1 w_2\, 
$$
$$ 
    w_0+w_1+w_2+w \, \in \, \S \ .
$$ 
As we have seen,
these conditions determine $w_1\tpbar\,w_2$ uniquely 
in the generic case where
$w_1\notin W_2$ and $w_2\notin W_1$.  
In the special cases where this is not true, 
a scaling ambiguity remains.
To supplement (16) for such cases,
we can stipulate
that $w_0\tpbar w_0=w_0$, 
and more generally 
that $w_0\tpbar w_2 = w_2$ 
and  $w_1\tpbar w_0 = w_1$.
These rules\footnote{$^\dagger$}
{
 In the previous section we already introduced rules for the special
 cases; the rules stated here are simply their instances for the
 situation at hand.  If we have restated them here, it is only in order to make the
 definition of $\tpbar$ more self-contained.
 }
%
%
render $w_1\tpbar\,w_2$ unique.
For example
if $w_1\in W_2$ then $w_1\in W_1\cap W_2$, whence $w_1=\lambda w_0$
since, as always, $\dim(W_1\cap W_2)=1$.
Therefore $w_1\tpbar\,w_2= (\lambda w_0)\tpbar\,w_2=\lambda w_2\,$,
exactly as in (11).


We learned in the previous section 
[following eq. (15)]
that $\tpbar$ 
would be
bilinear when
defined in this manner.\footnote{$^\flat$}
{One can also deduce the bilinearity of $\tpbar$ directly from the
 definition (16), if one proves first the following useful
 lemma: The set of first rows $(a,b)$ which make a square with a fixed
 second row $(c,d)$ is closed under addition and scalar multiplication;
 and similarly for columns instead of rows.  Closure under scalar
 multiplication we already noticed, and closure under sum can be deduced
 from the general square-form (15). Taken together,
 the row and column assertions suffice to prove linearity of $\tpbar$ in
 both arguments.}
Therefore 
(compare definition (2) of tensor-product in Section~2)
it induces a linear map $\Phi:W_1{\otimes}W_2\to V$.  
In fact $\Phi$ is an isomorphism.  
To see this, let's go back to the
representation of $V$ as $V_1 {\otimes} V_2$, and write 
$w_0=\alfa_0 {\otimes} \beta_0$, 
$w_1=\alfa {\otimes}\beta_0$,
$w_2=\alfa_0 {\otimes} \beta$.
Then 
as one sees by comparing 
(15) with (16),
$w_1\tpbar w_2=\alfa {\otimes}\beta$.  This
means first of all that 
the simple vectors $\alfa {\otimes}\beta\in\S$ 
coincide with the vectors of
the form $w_1\tpbar w_2$ for some $w_i\in W_i\ (i=1,2)$.
%
%
Consequently, we can build up
a basis of $V$ by choosing a basis
$\SetOf{e_j}{j=1\cdots\dim W_1}$ for $W_1$,
and a similar basis
$\SetOf{f_k}{k=1\cdots\dim W_2}$ for $W_2$,
and then taking our basis-elements to be 
$e_{jk}=e_j\tpbar\,f_k$.
That these $e_{jk}$ constitute a basis for $V$
follows from the fact that the $e_j$ (respectively the $f_k$)
have the form $\alfa_j {\otimes} \beta_0$ (resp. $\alfa_0 {\otimes}\beta_k$),
whereby 
the $\alfa_j$ (resp. $\beta_k$) constitute a basis of $V_1$ (resp. $V_2$)
if and only if the $e_j$ (resp.  $f_k$)
constitute a basis of $W_1$ (resp. $W_2$),
and furthermore
$e_{jk}=e_j\tpbar\,f_k=\alfa_j{\otimes}\beta_k$.

The upshot is that a ``special basis'' for $V$
(i.e. a basis of vectors $\alfa_j {\otimes}\beta_k$)
is the same thing as a pair of bases for $W_1$ and $W_2$, 
modulo the familiar $GL(1)$ ambiguity that one can rescale the $W_1$-basis
by $\lambda$ if one simultaneously rescales the $W_2$-basis by $1/\lambda$.
Recall now 
from Section~2
that our ``second possible answer'' to what constitutes a
tensor-product structure for $V$ was 
``an equivalence-class $\T$ of special bases for $V$''.
We have thus demonstrated that from $\S\subseteq V$ one can derive
uniquely a tensor-product structure in that sense.
%
Conversely, given such a structure $\T$, 
we immediately obtain $\S$ from it as
the union of all of the members of the special bases that comprise $\T$.
To the extent that $\S$ is a simpler and more natural object than an
equivalence class of special bases
(and is also more intrinsic to $V$), 
we have  reason to maintain that 
in $\S$ we have an answer to
the question, ``What does it mean for  $V$ to be a tensor
product?''.

\subsection {Using ``pointed vector spaces''}  
The above construction began with an arbitrarily chosen ``base-vector''
$w_0\in V$ such that $W_1=\pi_1 w_0$ and $W_2=\pi_2 w_0$.  
The ambiguity inherent in such a choice
does not 
impugn 
our demonstration of the equivalence, 
$\T \leftrightarrow \S$, 
but it does mean that in the procession,
$(V_1,V_2) \,\to\, V\hbox{-cum-}\S \,\to\, (W_1,W_2)$,
a different choice of $w_0$ would produce a different pair of spaces,
$W_1$, $W_2$. 
If desired,  
one could arrange for $W_1$ and $W_2$ to be unique 
by working with ``pointed vector spaces'', 
i.e. by equipping $V_1$ and $V_2$ with distinguished ``base-points'', 
$\alfa_0\in V_1$ and 
$\beta_0\in V_2$, 
and then taking
$v_0=\alfa_0{\otimes}\beta_0\in V_1{\otimes}V_2$ 
to be the base-point of $V$.  
Our construction above (with $w_0$ taken to be $v_0$)
would then recover the pairs
($V_1,\alfa_0$) and 
($V_2,\beta_0$) 
essentially uniquely from the triple ($V,\S,v_0$).

\smallskip
\REMARK  
Interestingly, the ``histories Hilbert spaces'' $\H$ that play a role
in Quantum Measure Theory [3] automatically come with distinguished
vectors $|\Omega\rangle\in\H$, where $\Omega$ represents the full
history-space (the unit of the corresponding event-algebra).
However it is generally false for coupled subsystems 
that $\H$ for the composite system 
is the Hilbert-space tensor product 
of the $\H$'s for the subsystems.
(Even when the vectorspace dimensions match, the norms in general will not.)
%

%
\section{6.~Categorical matters (and a shortcoming of the functor concept)}
From a given vector space $V$ one can form new spaces, like the
dual-space  $V^*$ or the double dual $V^{**}$.  With two vector spaces, there
are other possibilities, including their direct sum, their tensor product,
and so forth.  Although a vector space formed in one of these ways will
be isomorphic to infinitely many other vector spaces, its ``inner
constitution'' will in general be distinctive, with the result that it
will to a certain extent ``remember where it came from''.  One may say
then that it {\it\/carries the structure of\/} a dual space, a direct
sum, or a tensor product, as the case may be.  In each instance one can try
to identify concretely where this extra information resides, and for 
a vector-space $V$ that arose as a tensor product, our discussion
has pointed to the set $\S$ of simple vectors within $V$ as the
pertinent structure.
Adopting a notation that keeps track of $\S$,
we may say
that from an ordered pair $(V_1,V_2)$ of vector spaces, there arises
via tensor-product the ordered pair $(V,\S)$.

A question then is to what extent the transformation
$(V_1,V_2)\to(V,\S)$ is reversible.  How perfectly does $V$ remember
where it came from, or to ask this another way, how well can we
reconstruct $V_1$ and $V_2$, given $V$ and $\S$?  
When we dealt with pointed
spaces, we discovered that $(V_1,V_2)$ could ``in essence'' be recovered
fully.  But in the unpointed case, it appeared that although
$(V,\S)$ is determined by $(V_1,V_2)$, the latter could be
recovered from the former only up to some sort of $GL(1)$ ambiguity.
This suggests that in the pointed case a vector space carrying the
structure of a tensor product is in some sense equivalent to the factor
spaces from which it arose, whereas in the un-pointed case there is only
partial equivalence.

But what concept of equivalence is implicitly animating these
expectations?  Simple isomorphism will not do, being
too narrow in one way 
(because by definition structures of different types like $(V_1,V_2)$ and $(V,\S)$ cannot be isomorphic)
%
and too broad in another way (because, for example, any two vector
spaces of equal dimension are isomorphic).
Maybe one can put the underlying thought into words by saying that ``$A$
and $B$ are equivalent if $B$ can be constructed from $A$ and vice
versa.''
I am not sure that mathematics knows any 
framework
which really does
justice to this thought, but perhaps the category-theoretical concepts
of functor and natural isomorphism come closest to 
providing one,
and so it
seems worth considering how they apply to the question at hand.
%
%
We will take this up momentarily, but first let's see a very simple
illustration of how $\S$ is able to remember ``where $V$ came from''.

\subsection {A small illustration: topology remembers spin}  
As a small illustration of how the set $\S$ of simple tensors encodes
the structure of $V$ as a tensor product, let us return to the example
of 
$spin$-$3/2$ ${\otimes}$ $spin$-$1$ versus
$spin$-$1/2$ ${\otimes}$ $spin$-$5/2$.
To distinguish these two possible provenances of $V$,  
one from the other, 
it is enough to pay attention to the topology of $\S$,
for example its dimensionality.
Taking into account that an element of $\S$ has by definition the form
$\alfa\tensor\beta$, and that $\alfa$ and $\beta$ are unique modulo the
obvious $GL(1)$ ambiguity, we can observe that the (complex) dimensionality
of $\S$ is one less than the sum of the dimensionalities of the factor
spaces.  In our examples this yields for $\dim(\S)$
the  respective values,
$4+3-1=6$ for {\bf 3/2}$\,\otimes\,${\bf 1}
and $2+6-1=7$ for {\bf 1/2}$\,\otimes\,${\bf 5/2}.
In fact, it is easy to verify that this simple test works in general.
If we know that $V$ arose from the combination of two spins, then the
topological dimension of $\S$ determines fully what those spins were.
Of course (and as we have now seen in great detail) 
the same information can be deduced with a bit more work from
the dimensionalities of the maximal linear subspaces of $\S$, or from
the dimensionalities of $\M_1$ and $\M_2$ as ``foliations'' of $\S$. 

\subsection {A functorial gloss on our constructions}
Now back to categories, functors, and natural isomorphisms.  
A {\it\/category\/} is basically a
collection of spaces of a given type (its ``objects'') and of
structure-preserving mappings between these spaces (its ``morphisms'').
For our purposes it will be best to limit the 
latter
 to isomorphisms,
i.e.~to require them to be invertible.
A {\it\/functor\/} between two categories, I and II, is a kind of black box
that converts the objects and morphisms of category-I to objects and
morphisms of category-II while preserving composition of morphisms.
Conceptually, it is telling you that you can build
spaces and mappings of type II from spaces and mappings of type I
(but unfortunately it is not telling you {\it\/how\/} to do so.)


The two categories of interest to us here can be denoted
as $\VEC\times\VEC$ and $\TVEC$, where the former is the category of
pairs $(V_1,V_2)$  and the latter\footnote{$^\star$}
{The ``T'' in $\TVEC$ is meant to suggest the word ``tensor''}
is the category of pairs $(V,\S_V)$.
A morphism in $\VEC\times\VEC$ will thus be a pair of linear
isomorphisms, while a morphism in $\TVEC$ will be a linear isomorphism
between vector spaces that preserves their respective subsets $\S$.
When our spaces are pointed, all these isomorphisms will of course also
need to preserve the respective base-points.
Let us now describe some of our constructions in terms of functors between
$\VEC\times\VEC$ and $\TVEC$,
concentrating for the time being exclusively on the pointed case.  


The first
functor of interest, 
which we will designate as \hbox{${\otimes}:\VEC\times\VEC\to\TVEC$}, 
is that induced by the tensor product itself.  
It takes a pair of vector spaces $(A,B)$ to their tensor
product, $V=A{\otimes}B$ 
equipped with its space $\S_V$ of simple vectors
$\alfa{\otimes}\beta$,
and it takes a pair $(f,g)$ of (invertible) linear functions between
vector spaces to their tensor product $f{\otimes}g$.  
Conversely, given an object $(V,\S_V,v_0)\in\TVEC$ 
(where I've now indicated the base-point $v_0$ explicitly), 
we saw how to locate within $V$ 
the subspaces $W_1=\pi_1(v_0)$ and $W_2=\pi_2(v_0)$, 
which were certain maximal linear subsets of $\S_V$.  
Thereby, we in effect defined a second functor,
$D:\TVEC\to\VEC\times\VEC$, 
that goes in the direction opposite to ${\otimes}$,
and for which $D(V,\S_V,v_0) = ((W_1,v_0),(W_2,v_0))$.  
%
Of course one has not defined a functor fully until one tells
how it acts on morphisms, but that is self-evident for $D$.
A morphism in $\TVEC$
from $(V,\S_V,v_0)$ to $(V',\S_V',v_0')$ is nothing but an
invertible linear function, $f:V\to V'$, 
such that $f[\S_V]=\S_V'$ and $f(v_0)=v_0'$.
Such an $f$ induces immediately a pair of
(basepoint preserving)
functions $f_1:W_1{\to}W_1'$ and $f_2:W_2{\to}W_2'$,
and so $D(f)=(f_1,f_2)$.

Now 
what of
the expectation that ${\otimes}$
and $D$ are in essence each other's inverses?  Were that literally true,
we would be able to express it 
by writing
${\otimes}\compose D=1$ and $D\compose{\otimes}=1$,
but unfortunately both equations are,
strictly speaking, false.
Consider first the composed functor, $D\compose{\otimes}$.
What happens when we apply it to the pair of pointed vector spaces 
$((A,\alfa_0),(B,\beta_0))$? 
Tracing through the definitions, we find
${\otimes}((A,\alfa_0),(B,\beta_0))=(A {\otimes} B, \S_{A{\otimes}B}, \alfa_0 {\otimes} \beta_0)$, 
and then 
$
  D(A {\otimes} B, \S_{A{\otimes}B}, \alfa_0 {\otimes} \beta_0)
  = ((W_1,w_1),(W_2,w_2))
$,
where
$W_1=\pi_1(\alfa_0{\otimes}\beta_0)=A {\otimes}\beta_0$,
$W_2=\pi_2(\alfa_0{\otimes}\beta_0)=\alfa_0{\otimes}B$,
and $w_1=w_2=\alfa_0{\otimes}\beta_0$.
In other words,
$$
   (D\compose{\otimes})((A,\alfa_0),(B,\beta_0)) = ((A {\otimes}\beta_0, \alfa_0{\otimes}\beta_0), (\alfa_0{\otimes}B, \alfa_0{\otimes}\beta_0))
  \eqno(17)
$$
While $(D\compose{\otimes})((A,\alfa_0),(B,\beta_0))$ is thus not exactly identical with $((A,\alfa_0),(B,\beta_0))$,
there is between them an obvious correspondence,  
$((A,\alfa_0),(B,\beta_0)) \longleftrightarrow (D\compose{\otimes})((A,\alfa_0),(B,\beta_0))$,
%
%
given by the linear isomorphisms,
$$
    \alfa \, \longleftrightarrow \, \alfa\,\otimes\beta_0  
    \qquad \hbox{and} \qquad
    \beta \, \longleftrightarrow \, \alfa_0\,\otimes\beta  \eqno(18)
$$
The bijection (18) is an instance of what is called a
{\it\/natural isomorphism\/} between functors, 
and so 
category theory gives us a precise way to express that $D$
is effectively a right-inverse of ${\otimes}$ by saying that
${\otimes}\compose D$ is ``naturally isomorphic'' to the
identity-functor, a relationship which we will write as
$$
    {\otimes} \compose D \NatIsom 1 \ .  \eqno(19)
$$

It is evident from its definition that
the correspondence (18)
establishes an isomorphism
between two {\it\/objects\/}  in $\VEC\times\VEC$.
If read from left to right, it is a mapping
 $\Psi:((A,\alfa_0),(B,\beta_0))\to(D\compose\otimes)((A,\alfa_0),(B,\beta_0))$, 
but
what is it
that earns $\Psi$ the title, ``natural'',
thereby authorizing the use of the symbol $\NatIsom$ in (19)?
%
%
It is that 
$\Psi$ also
induces the correct
correspondence between {\it\/morphisms\/}
by converting
$(f_1,f_2)$ into $(D\compose\otimes)(f_1,f_2)$.
This is
self-evident when one unpacks the definitions (cf. (20) below),
but
even without unpacking the definitions, we could have been
assured that $\Psi$ was natural, if we had reflected
that it was defined {\it\/intrinsically\/},
utilizing
nothing more than the structures displayed in (17). 
Indeed, I think it would be fair to say that
this possibility of being constructed from intrinsic information 
without the intervention of any arbitrary choices 
is what best expresses the intuitive meaning of ``naturality''.

The distinction between plain isomorphism $\isom$ and natural
isomorphism $\NatIsom$ is perhaps most familiar in the example of dual
vector-spaces, where both $V^{**}$ and $V^*$ are isomorphic to $V$, but
only the isomorphism between $V^{**}$ and $V$ is natural.  Given an
element $v\in V$ one can define $v^{**}\in V^{**}$ by the equation, 
$v^{**}(f)=f(v)$, where $f\in V^*$.  On the other hand, there is no way
to pass deterministically from $v$ to an element $f\in V^*$ without the
aid of a basis for $V$, or a metric, or some such auxiliary structure.

\medskip
\REMARK  A natural isomorphism sets up an equivalence between functors
in much the same way 
as a similarity transformation sets up an equivalence
between group representations.  If 
$R_1$ and $R_2$ are representations of the group $G$ 
related by the similarity transformation $S$, then
$S R_1(g) S^{-1} = R_2(g)$, or equivalently $S R_1(g)=R_2(g)S$.
Now 
if 
we replace $R_1$ and $R_2$ by functors
$F_1$ and $F_2$, and the arbitrary group-element $g$ by an arbitrary
morphism $f$, 
we obtain the condition 
for a family of invertible morphisms $S$ 
to define a natural isomorphism between $F_1$ and $F_2$, 
namely $S F_1(f)=F_2(f)S$.
Often this last equation is represented by drawing the commutative diagram,
$$
         \matrix{F_1X & {F_1f \atop \rightarrow} & F_1Y \cr
                 S_X  \downarrow\quad  &   &  \quad \downarrow S_Y \cr
                F_2X & {\rightarrow  \atop F_2f } & F_2Y \cr}   
$$
where $X$ and $Y$ are any objects in the category 
and $f:X\to Y$ is any morphism
between them.
When, as in our case, 
$F_1$ is the identity functor, 
the diagram for $S:1\to F$
simplifies to 
$$
         \matrix{X & {f \atop \rightarrow} & Y \cr
                 S \downarrow\quad  &   & \quad \downarrow S \cr
                FX & {\rightarrow  \atop Ff } & FY \cr}   \eqno(20)
$$
%
One sees in this case that the functor $F$ must
be a bijection between 
the morphisms $f$ and the morphisms $Ff$;
and conversely, 
the fact
that $F$ is such a bijection
captures to a large extent everything that
the equation $F\NatIsom1$ means. 

Having established that
$D\compose{\otimes}\NatIsom1$,
let us now 
try to demonstrate
the complementary equivalence,
${\otimes}\compose D\NatIsom1$.
Following the same steps as before, 
let us apply the functor
${\otimes}\compose D$  to the object $(V,\S_V,v_0)\in\TVEC$,
obtaining first
$D(V,\S_V,v_0)=((W_1,v_0),(W_2,v_0))$
and then
${\otimes}((W_1,v_0),(W_2,v_0))=(W_1{\otimes}W_2,\S_{W_1{\otimes}W_2},v_0{\otimes}v_0)$,
which taken together tell us that
$$
    ({\otimes}\compose D) (V,\S_V,v_0) = (W_1{\otimes}W_2,\S_{W_1{\otimes}W_2},v_0{\otimes}v_0)
  \eqno(21)
$$
Can we exhibit a natural isomorphism equating
$(W_1{\otimes}W_2,\S_{W_1{\otimes}W_2},v_0{\otimes}v_0)$ 
to $(V,\S_V,v_0)$,
and therefore ${\otimes}\compose D$ to the identity functor?
To this question we already have the answer
in the form of
the isomorphism, $\Phi:W_1{\otimes}W_2\to V$,
which we constructed earlier with the aid of the 
intrinsically defined
product $\tpbar$,
and for which $\Phi(w_1{\otimes}w_2)=(w_1\tpbar w_2)$.
As with $\Psi$ before, 
it is straightforward to verify that  $\Phi$ is natural, 
as indeed it had to be, given its intrinsic nature.
Therefore ${\otimes}\compose D\NatIsom1$.
%

We have now proven that the composition of ${\otimes}$ with
$D$ in either order is naturally isomorphic to the identity.
Thus category theory, by introducing the concept of natural isomorphism
$\NatIsom$ as a replacement for strict equality, has given us a
way to make precise (and then to verify) the informal claims that, in
the pointed case, the functor ${\otimes}$ is invertible and that $D$ is
its inverse.

Turn now to the unpointed case and to our expectation that it will not
be possible to recover the pair $(A,B)$ from $(A{\otimes}B,\S)$ when the
spaces involved are not equipped with base-points.
Can we also corroborate this expectation within the categorical framework?
Stated formally, the question is
whether there exists a functor
$D:\TVEC\to\VEC\times\VEC$ which is a ``left inverse'' to ${\otimes}$ in
the sense that $D\compose{\otimes}\NatIsom1$.  
In fact, it's easy to see
that no such functor can exist.  
Were $D$ such a functor then, as we
noticed in connection with (20), 
the mapping,
 $f\mapsto (D\compose\otimes)f$,
would have to be invertible
for morphisms,
$f:(A,B)\to(A',B')$,
of the category $\VEC\times\VEC$,
where $f$ is
by definition 
a pair $(g,h)$ of individual morphisms in $\VEC$.
This, however, is
clearly impossible because 
the functor ${\otimes}$ 
(and therefore its composition with $D$ if the latter existed)
fails to be injective, 
since it
maps both  $f=(g,h)$ and $\tilde{f}=(\lambda g,h/\lambda)$
to the single morphism $g{\otimes}h=(\lambda g){\otimes}(h/\lambda)$.
In other words,
${\otimes}$ acting on morphisms
is not injective but many-to-one, 
the ``many'' being parametrized by a non-zero scalar $\lambda$ which 
embodies the same $GL(1)$ ambiguity we met with earlier.
This confirms that the equation $D\compose{\otimes}\NatIsom1$ 
can have
no solution, and a fortiori that
the functor ${\otimes}$ is not invertible.

A somewhat simpler example of the same nature occurs in connection with
the attempt to represent a spinor geometrically.  Starting from a
2-component Weyl spinor $\zeta$, for example, one can derive
algebraically a so-called null flag $F$, which consists of a lightlike
vector together with a half-plane matched to the vector.  [4]
But because
vectors are quadratically related to spinors, both $\zeta$ and $-\zeta$
give rise to the same flag $F$, whence one can recover 
the spinor from the flag
only up to an unknown sign.  
(This loss of information was inevitable, because spinors change sign after
rotating through $2\pi$, whereas  vectors do not.)
To couch these relationships in categorical language, one could
introduce a category of spinor-spaces and a category of spaces of
null-flags and a functor $\phi$ from the former to the latter.  Like
${\otimes}$ above, $\phi$ would not be invertible, because it would be
$2\to1$ on morphisms.
At best, one might be able to devise, as a kind of {\it\/right\/} inverse to $\phi$, 
a ``functor manqu{\'e}'' or ``functor up to sign'' going from flag-spaces 
to spinor spaces.
Its existence would proclaim that, although not fully a geometrical
object, a spinor is nevertheless ``geometrical up to sign''.

\medskip\REMARK  Despite its utility, the concept of functor 
does not necessarily illuminate 
the connection between its inputs and its outputs 
as fully as one might have expected it to do,
because unlike a morphism, it is blind to the individual elements of the
spaces on which it acts; by definition it does not ``look inside''.
Thus if $\phi$ is a functor 
and $X$ a space (or a mapping), 
and if $Y$ is the space (or mapping) that results when $\phi$ acts on $X$,
then the equation $Y=\phi(X)$
tells us {\it\/that\/} $Y$ is in some sense built from $X$,
but it tells us nothing concretely about {\it\/how\/}  $Y$ is built from $X$.\footnote{$^\dagger$}
{Could Bourbaki's concept [5] of ``deduction of structures'' come any closer to doing this?}
%
%
For example
$X$ could be a spinor-space,
and $\phi$ 
the above functor.
Then $Y=\phi(X)$ would be the space
comprised of all the null flags derived from the spinors comprising $X$.
But if $\zeta\in X$ were an {\it\/individual\/}
spinor in $X$,
and if $F$ were the individual flag derived from $\zeta$,
the rules governing functors would not allow us to write ``$F=\phi(\zeta)$'',
even though it might seem natural to do so, and 
even though we know perfectly well what we would mean by it!
%


\section{7.~Questions; further developments; connection to quantum field theory}  
In conclusion, let me mention a few questions and possible further
developments suggested by the above considerations.

The most important of the constructions introduced in Sections 4 and 5
revolve around the ``foliations'' $\M_1$ and $\M_2$, the corresponding
mappings $\pi_1$ and $\pi_2$, the {\it\/square\/} concept, and the product
$\tpbar$ which results from these via the definition (16).

An obvious question that one might ask is how these distinctive
ingredients generalize to the tensor product of three or more vector
spaces.  One could of course just treat a threefold product like
$A\tensor{}B\tensor C$ as an iterated pairwise product like
$(A{\tensor}B)\tensor C$, but a more symmetric construction ought to be
possible, and one might expect it to uncover some new structures that
are not visible in connection with simple pairwise products like
$A\tensor{}B$.

One might also wonder whether there was anything of
interest to be learned from the study of the various symmetry types that
become possible when two or more of the factor-spaces are equal to each
other.  
For example when $V\subseteq{A}\tensor A$ is the subspace of symmetric
tensors, two natural analogs of $\S$ as used above would be the set of
tensors of the form, $\alfa\tensor\beta+\beta\tensor\alfa$, or even more
simply, of the form $\alfa\tensor\alfa$.  
Or for the anti-symmetric tensor product of $A$ with itself, 
the set of tensors of the form, 
$\alfa\wedge\beta=\alfa\tensor\beta-\beta\tensor\alfa$, 
would be a natural analog of $\S$. 
To what extent, and in what
form, could one repeat the above discussion with one of these subsets
replacing $\S$?
And still more generally, what might be an analog of $\S$ belonging to
the non-abelian symmetry-types 
(those corresponding to more general Young tableaux)
that 
arise as subspaces of higher products,
${A}\tensor A \cdots \tensor A$,
and which 
are neither ``bosonic'' nor ``fermionic''?

Another sub-case of obvious interest is that where the vector spaces
are equipped with metrics, in particular where they are Hilbert spaces.
One might then expect orthonormality to play a role, but would any 
additional, unexpected features of interest show up?

Our discussion so far has proceeded in finite dimensions.  If we want to
generalize it to infinite-dimensional vector spaces, a whole raft of
further questions will appear, some of which concern the definition of
tensor product itself.
Clearly, the subset $\S$ of simple tensors $\alfa\tensor\beta$
within $V=A\tensor{B}$ can be defined without difficulty, but will our
constructions based on it also go through as before?  Will they still let us recover
$A$ and $B$, and will they still lead us as in Sections 4 and 6 (say in
the pointed case) to a functor $D$ inverse to $\tensor$?
In all of this, what consequences might flow from ambiguities in the
definition of $\tensor$?  When $A$ and $B$ are Hilbert spaces,
$A\tensor{B}$ qua Hilbert space is unambiguous, but when they are only
Banach spaces (normed vector spaces), many different spaces
$A\tensor{B}$ have been defined [2]. One may wonder then
whether $\S$ will still be able to ``remember'' which specific choice of
$\tensor$ went into the creation of $V$.


Among infinite-dimensional vector spaces, the Hilbert spaces have a
special significance for quantum theories.  Although the ambiguity in
defining $A\tensor B$ is not an issue when $A$ and $B$ are Hilbert spaces, 
it can happen 
in connection with quantum field theory
that the notion of tensor product itself seems to be transcended. 
If one divides a Cauchy surface into two complementary regions, then
naively one would expect the overall Hilbert space $\H$ of the
field-theory to be the tensor product of Hilbert spaces associated
with the two regions, just as happens with composite systems in ordinary
quantum mechanics.  Unfortunately, this would conflict with the fact
that the operator algebras associated with the two regions (technically
with their domains of dependence in spacetime) are known (for free
fields) to be of ``type III'', this being intimately linked to the
infinite entanglement-entropy between the two regions.  One still has
operator subalgebras for the regions (so-called coupled factors), but
these subalgebras cannot be interpreted as acting on the separate
factors of a tensor-product.  One thus confronts something like a
tensor product of operator-algebras that does not derive from a
tensor-product decomposition of the underlying Hilbert 
\hbox{space $\H$}.  
(Adopting the language of ``quantum systems'', one might say that one is
dealing with ``subsystems which possess observables but lack
state-vectors''.)\footnote{$^\flat$}
{This is not quite the same as saying that a type-III factor lacks pure
 states.  As most commonly defined in the theory of operator-algebras, a
 pure state on $\A$ is an extreme point in the convex set of normalized
 positive linear functionals on $\A$.  It is known that such pure states
 exist copiously, and one could thus entertain them as generalized
 state-vectors, since in finite dimensions, state-vector = pure state.
 However when $\A$ is a type-III factor, its pure states seem to be
 mathematically pathological (perhaps even ``ineffable''), and one could
 plausibly regard them as unphysical.  See
 [6].}



In the absence of a tensor-product structure for $\H$, 
the notion of simple-vector is not defined, 
and therefore neither is our subset $\S\subseteq\H$.
Nevertheless, one might hope that some 
generalization of simple vector,
and some corresponding subset of $\H$, 
could serve a similar function.
%
%
%
%
Simple vectors are tensors of rank 1, but the tensors of ranks, 2, 3, 4,
etc. also respond to the tensor-product structure of $V$.  
Could it be that suitable analogs of the spaces of such tensors 
(or better, of the tensors of finite ``co-rank'' in some suitable sense)
are able to capture the structure of coupled factors of types-III or II, 
and in so doing shed light on features like
the area-law for entanglement-entropy?
%
%
Especially salient in this
connection is 
the ``spatiotemporal cutoff'' needed
to render
the entropy finite [7].
Physically, such a cutoff needs to be frame-independent
(locally Lorentz invariant), and it seems suggestive that an analog of $\S$, if it could be defined, 
would not obviously need to refer to any arbitrarily chosen reference-frame.


As a first approach to some of these questions, one could ask in finite
dimensions how to relate the operator-algebra framework to that of the
present paper.  Indeed, one might have thought to identify a
tensor-product-structure for $V$, not with the family $\S$ of simple
vectors in $V$, but with a pair of commuting operator-subalgebras which
generate the algebra $L(V)$ of all linear operators on $V$ and which
have in common only the multiples of the identity-operator (like
Murray-von Neumann coupled factors but without any specialization to
complex numbers or self-adjointness).
%
%
The advantage of such an alternative approach would be that the algebras
$L(A)$ and $L(B)$ reappear bodily in $L(A\tensor B)$, whereas the spaces
$A$ and $B$ themselves need to be excavated from $A\tensor B$ more
painfully, as we have seen in great detail above.  (Quantally speaking,
the ``observables'' of a subsystem carry over to the composite system,
whereas the ``states'' do not.  But see the remark below.)
Its disadvantage would be that an algebra of operators in $V$ is a
considerably more complicated beast than the simple subset
$\S\subseteq{V}$.  Be that as it may, it's clear that the two viewpoints
are related.  For example, an operator acting only on $A$ (an operator
in $L(A)\tensor{\bf{1}}$), or only on $B$,  will automatically be an operator that
preserves $\S$, suggesting how one might derive $L(A)$ and $L(B)$ from
$\S$.

\smallskip
\REMARK In the context of Quantum Measure Theory  [3], 
the histories-hilbert-space associated to a subsystem actually does
reappear as a true subspace of the histories-hilbert-space of the full
system, the reason being that an {\it\/event\/} in a subsystem is
{\it\/ipso facto\/} an event in the full system.  Moreover this subspace
carries a distinguished ``base point'', as remarked in Section~5.  When,
in addition, the overall quantum-measure is the product-measure 
(as for ``non-interacting subsystems in a product-state''), 
the full
histories-hilbert-space is the tensor product of the subspaces, and the
aforementioned advantage of an approach via operator algebras
disappears.
%

Let us return, finally, to finite dimensions and to the cone $\S$ of
simple vectors within $V$, on which most of our work has been based.  We
have seen how $\S$ endows $V$ with the structure of a product space, but
we did not provide (or even ask for) a simple criterion that would let
us recognize whether a given subset $\S$ could actually play the role
assigned to it.  That is, we did not provide necessary and sufficient
conditions for there to exist an isomorphism mapping $V$ to a space
$A\tensor B$ that would map $\S$ to the set of tensors of the form
$\alfa\tensor\beta$.
One trivially adequate criterion is that the re-constructions undertaken
in Section 5 should succeed, and in particular that the building up of
the {\it\/squares\/} should never encounter an obstacle.  But one might wish for
criteria that were more self-contained and more simply stated.  Given
that the rays in $\S$ constitute a ``Segre variety'', one might hope
that the Algebraic-Geometry literature would contain something of this
sort.

Alternatively, rather than seeking axioms for $\S$, one might instead
seek axioms for the squares themselves, i.e. axioms for quadruples of
vectors in $V$.  A tensor-product structure for $V$ would then be a set
of quadruples satisfying these axioms.



\bigskip
\noindent
This research was supported in part by NSERC through grant RGPIN-418709-2012.
This research was supported in part by Perimeter Institute for
Theoretical Physics. Research at Perimeter Institute is supported
by the Government of Canada through Industry Canada and by the
Province of Ontario through the Ministry of Economic Development
and Innovation.  

\ReferencesBegin                             


\ref [1] 
For more detail on these definitions, see:
\sepref
Shlomo Sternberg, {\it\/Lectures on Differential Geometry\/} (Englewood Cliffs, N.J: Prentice-Hall, 1964)
\sepref
Tracy Y. Thomas, {\it\/Concepts from Tensor Analysis and Differential Geometry\/} 
(Academic Press, 1961), pages 7ff 
\sepref
Saunders Mac Lane, {\it\/Homology\/} (Academic Press, 1963), pages 138ff
\sepref
Chapter 1 of Sternberg contains examples of most of the definitions,
including implicitly the definition of $V\tensor W$ as the space of
bilinear mappings of $V^*\times W^*$ into $\Reals$.  The time-honoured
definition of a tensor in terms of transformation laws for its
components is presented in Thomas.


\ref [2] 
A. Grothendieck, ``Produits tensoriels topologiques et espaces nucl{\'e}aires'', 
\journaldata{Memoirs of the American Mathematical Society} {16} {} {1955}

\ref [3] 
Rafael D.~Sorkin,
``Quantum Mechanics as Quantum Measure Theory'',
   \journaldata{Mod. Phys. Lett.~A}{9 {\rm (No.~33)}}{3119-3127}{1994},
   \lbr
   \eprint{gr-qc/9401003},
   \lbr
   \eprint{http://www.pitp.ca/personal/rsorkin/some.papers/80.qmqmt.pdf}
\sepref
Fay Dowker and Rafael D. Sorkin, 
``An intrinsic causality principle in histories-based quantum theory: a proposal''
(to appear)
\sepref
  Fay Dowker, Steven Johnston, Sumati Surya, ``On extending the Quantum Measure'',
  \journaldata{J. Phys. A}{43}{505305}{2010}
  \eprint{arXiv:1007.2725 [gr-qc]}



\ref [4]  
Roger Penrose, ``Structure of space-time'', 
in Battelle Rencontres: 1967 Lectures in Mathematics and Physics 
(edited by Cecile M. DeWitt and John A. Wheeler)
(New York: Benjamin, 1968)

\ref [5] 
N. Bourbaki, {\it Theory of Sets } (Paris: Hermann, 1968),
Chapter IV

\ref [6] 
Bruce Blackadar (2017), {\it\/Operator Algebras: Theory of C*-Algebras and von Neumann Algebras\/}, 
\eprint{https://packpages.unr.edu/media/1224/cycr.pdf} 
\lbr  
See III.2.2.3 and the remarks in III.2.2.15

%

\ref [7] 
Rafael D. Sorkin, 
``On the Entropy of the Vacuum Outside a Horizon'',
  in B. Bertotti, F. de Felice and A. Pascolini (eds.),
  {\it Tenth International Conference on General Relativity and Gravitation (held Padova, 4-9 July, 1983), Contributed Papers}, 
  vol. II, pp. 734-736
  (Roma, Consiglio Nazionale Delle Ricerche, 1983),
  \lbr
  \eprint{http://www.pitp.ca/personal/rsorkin/some.papers/31.padova.entropy.pdf}
  \lbr
  \eprint{http://arxiv.org/abs/1402.3589}
\sepref
  Rafael D.~Sorkin,
``The Statistical Mechanics of Black Hole Thermodynamics'',
  in R.M. Wald (ed.) {\it Black Holes and Relativistic Stars}, 
  (U. of Chicago Press, 1998), pp. 177-194
  \eprint{gr-qc/9705006}\lbr
  \eprint{http://www.pitp.ca/personal/rsorkin/some.papers/92.chandra.pdf}

\end 

(prog1 'now-outlining
  (Outline* 
     "\f"                   1
      "
      "
      "
      "\\Abstract"          1
      "\\section"           1
      "\\subsection"        2
      "\\appendix"          1       ; still needed?
      "\\ReferencesBegin"   1
      "
      "\\ref "              2
      "\\end